# Electromagnetic properties and electronic structure of iron-based layered superconductor LaOFeP

Yoichi Kamihara,[1] Masahiro Hirano,[1,2] Hiroshi Yanagi,[3] Toshio Kamiya,[1,3] Yuji Saitoh,[4] Eiji Ikenaga,[5] Keisuke Kobayashi,[5,6] and Hideo Hosono[1–3]

[1]*ERATO-SORST, JST, in Frontier Research Center, Tokyo Institute of Technology, Mail Box S2-13, 4259 Nagatsuta, Midori-ku, Yokohama 226-8503, Japan*

[2]*Frontier Research Center, Tokyo Institute of Technology, Mail Box S2-13, 4259 Nagatsuta, Midori-ku, Yokohama 226-8503, Japan*

[3]*Materials and Structures Laboratory, Tokyo Institute of Technology, Mail Box R3-1, 4259 Nagatsuta, Midori-ku, Yokohama 226-8503, Japan*

[4]*Synchrotron Radiation Research Center, Japan Atomic Energy Agency, Kouto 1-1-1, Mikazuki, Sayo, Hyogo 679-5148, Japan*

[5]*JASRI/SPring-8, Kouto 1-1-1, Sayo-cho, Sayo-gun, Hyogo 679-5198, Japan*

[6]*NIMS/SPring-8, Kouto 1-1-1, Sayo-cho, Sayo-gun, Hyogo 679-5198, Japan*

**Abstract**

Structural, electronic, and magnetic properties of undoped and aliovalent-ion (Ca, F)-doped LaOFeP, which undergo superconducting transitions at transition temperatures ($T_c$) 4–7 K [Kamihara et al, J. Am. Chem. Soc. **128**, 10012, (2006)], were investigated. $T_c$ of the samples varied from 2.4 to 5.5 K in the undoped samples, and was increased up to ~7 K by Ca- and F-doping. The $T_c$ increases are correlated with a decrease in the lattice volume. LaOFeP exhibits paramagnetism in the normal conducting state. Photoemission spectroscopy combined with first-principle band calculations clarified that Fe 3$d$ ($d_{z2}$ + ($d_{xz}$, $d_{yz}$)) orbitals hybridized with P 3$p$ to form a Fermi surface. The band calculations also suggest that the 3d electron of the Fe in LaOFeP is basically in the low-spin configuration, and that the spin moment of LaOFeP is almost quenched, leading to the paramagnetism of the itinerant electrons.

# I. INTRODUCTION

In addition to various types of copper-based oxide superconductors, [1,2] several superconductors that contain a non-copper transition metal element as a major constituent, such as $Na_xCoO_2\cdot 1.3H_2O$, $Na_xV_2O_5$, $Cu_xTiSe_2$, and $Ln Ni_2B_2C$ ($Ln$ = Y, Tm, Er, Ho, or Lu), have recently been reported. [3–6] Moreover, a considerable number of iron-based compounds have been reported as exhibiting superconductivity, including inter-metallic compounds ($U_6Fe$, $Th_7Fe$ [7,8]), iron silicates ($R_2Fe_3Si_5$, $R$ = Sc, Y, Lu, and Tm [9,10]), and rare-earth-filled skutterudites ($Ln Fe_4P_{12}$, $Ln$ = La, Y [11,12]) whose transition temperatures ($T_c$) range from 1.8 to 7 K. All of these compounds show Pauli paramagnetic behavior in the normal conducting states, indicating that the magnetic moments of the irons are quenched. The quench of the magnetic moment is also observed in a high-pressure phase of elementary iron $\varepsilon$-Fe, which shows a superconducting transition at ~2 K. [13] We have recently found that LaOFeP exhibits metallic conduction near room temperature and undergoes a superconducting transition at ~4 K, [14] providing a new compound system to the iron-based superconductors.

LaOFeP is a member of the $LnOT_MPn$ family, where $Ln$ represents a 4$f$ rare-earth element, $T_M$ a transition metal element with a more than half-filled 3$d$ shell, and $Pn$ a pnicogen element. These compounds belong to the tetragonal $P4/nmm$ space group, [15–18] and the unit cell containing two chemical formula units is represented by $(La_2O_2)(Fe_2P_2)$. This crystal has a layered structure composed of an alternate stack of LaO and FeP layers (Fig. 1). A Fe ion is coordinated with four P ions to form a distorted $FeP_4$ tetrahedron, which is linked with neighboring $FeP_4$ tetrahedra in an edge-sharing manner to comprise the FeP layer. From the facts that iron monophosphide (FeP) exhibits metallic conduction [19] and the Cu$Ch$ layers in LaOCu$Ch$ ($Ch$ = S, Se), which are isostructural compounds to LaOFeP, form hole transport paths [20,21], we expected that the FeP layer in LaOFeP would also form a carrier conduction path. Such a structure, in which the carrier conduction layer is sandwiched by the wide-gap LaO layers, may be regarded as a multiple-quantum well (MQW) embedded in the crystal structure. [20] Therefore, strong electron–electron interaction is expected, owing to the two-dimensional carrier confinement in the FeP layers. In addition, the composition of LaOFeP tends to deviate from the stoichiometry primarily because it contains two different anions with different charge state. This non-stoichiometry involving O/P vacancies and/or anti-sites may also alter the carrier density around the Fermi surface, possibly inducing a significant change in $T_c$. Further, aliovalent ion doping may be more effective in altering the density of states responsible for the superconductivity. It is worth noting that the electromagnetic properties of LaO$T_M$P drastically change with the number of 3d electrons, i.e., an antiferromagnetic semiconductor for Mn (3d$^5$) [22], a ferromagnetic metal for Co (3d$^7$) [22], a superconductor for Ni (3d$^8$) [23], and a semiconductor for Zn (3d$^{10}$) [24]. It should be noted that a family compound, LaOFeAs, exhibits superconductivity at much higher temperatures up to 26 K upon F doping to the oxygen ion sites. [25] $T_c$ of the F-doped LaOFeAs shifts up to 43 K upon applying high pressure. [26] It makes the $Ln$O$T_M$$Pn$ family recognized as a new high $T_c$ superconductors system. After these publications, many preprint papers on $Ln$OFeAs ($Ln$ = lanthanide ions) have been reported during the review process of this paper. [27]

In this study, we have performed X-ray diffraction, electrical resistivity, and magnetic susceptibility measurements on LaOFeP, La(O$_{0.94}$F$_{0.06}$)FeP, and (La$_{0.9}$Ca$_{0.1}$)OFeP. It was clarified that $T_c$ of undoped LaOFeP ranged from 2.4 to 5.5 K, which correlated with the lattice parameters. Moreover, it was observed that the F-ion doping increased $T_c$ up to 7 K. Further, electronic and magnetic structures were examined by photoemission spectroscopy (PES) with the aid of first-principles band calculations. Here, we employed soft X-ray PES (SXPES) with an excitation energy of 701.24 eV and hard X-ray PES (HXPES) with a higher excitation energy of 7936.06 eV, which may make it possible to assign origins of PES signals because a photo-ionization cross-section of electron depends on the photon energy and atomic orbital of electron.

## II. EXPERIMENTAL

Several samples were prepared for nominally undoped, F-doped, and Ca-doped LaOFeP. All the samples were polycrystalline and prepared via the following solid-state reactions. Zimmer *et al.* reported a synthesis route for LaOFeP. [15] However, different routes were employed in this study to obtain higher-purity samples. We used dehydrated $La_2O_3$ and a mixture of compounds, which is composed of LaP, $Fe_2P$, and FeP (LaP-$Fe_2P$-FeP powder), as starting materials. To obtain the LaP-$Fe_2P$-FeP powder, La (Shinetsu Chemical; La with purity 99.5 mol%, containing Ce 0.33 mol%, Pr 0.01 mol%, Nd 0.06 mol%, Fe 0.22 mol%, etc.), Fe (Kojundo Chemical; >99.9%), and P (Rare Metallic Chemical; 99.999%) were mixed in a ratio of 1:3:3 and heated at 700 °C for 10 h in an evacuated silica tube. $La_2O_3$ was dehydrated by heating commercial $La_2O_3$ powder (Kojundo Chemical; 99.99%) at 600 °C for 10 h in air. Then, a 1:1 mixture of LaP-$Fe_2P$-FeP powder and the $La_2O_3$ powder was heated in an evacuated silica tube at 1200 °C for 40 h to prepare undoped LaOFeP. F-doping was performed by replacing a part of $La_2O_3$ with a 1:1 mixture of $LaF_3$ (Morita Chemical; 99%) and La metal in the starting materials for undoped LaOFeP. Ca-doped LaOFeP was synthesized by heating a mixture of FeP (prepared by heating a 1:1 mixture of Fe and P at 1000 °C for 10 h in an evacuated silica tube), dehydrated CaO (prepared by heating commercial $CaCO_3$ powder (Kojundo Chemical; 99.99%) at 920 °C for 10 h in air), and the starting materials for undoped LaOFeP in an evacuated silica tube at 1250 °C for 40 h. For the final reactions, the silica tubes were filled with high-purity Ar gas under 0.2 atmospheric pressure at room temperature to prevent collapse of the silica tubes. The Ar pressure in the tube was expected to prevent the evaporation of volatile components in the mixture, resulting in the formation of the near-stoichiometric LaOFeP.

All the samples were subjected to powder X-ray diffraction (XRD) and electrical resistivity measurements. Three samples (one of each was chosen from the undoped, the F-doped, and the Ca-doped samples, denoted as "representative samples" hereafter) were used for the other measurements. The crystallographic phases were characterized by XRD (Rigaku RINT-2500) using Cu Kα radiation. Lattice parameters were obtained by a least-squares fitting method using the diffraction angles collected in the 2θ range from 30 to 120°, and corrected them with an external reference method using a reference-grade Si powder (NIST SRM 640c).

Electrical resistivity measurements were conducted at 2–300 K using a DC four-probe technique. Magnetization was measured on the representative samples using a Quantum Design Physical Properties Measurement System (PPMS) with a vibrating sample magnetometer (VSM) option at 2–370 K. Core-level and valence-band structures were measured by HXPES (hν = 7936.06 eV; total energy resolution Δ$E$, which was estimated from the Fermi-edge broadening of an Au reference, was less than 250 meV) and SXPES (hν = 701.24 eV; Δ$E$ was approximately 130 meV) at 20 K on the undoped LaOFeP used for the magnetization measurements. These spectra were respectively measured at BL29XU [28-30] and BL23SU [31] beam-lines of SPring-8. Spin-polarized first-principles band calculations were performed based on the linearized augmented plane wave plus local orbitals (L/APW+lo) method at the density-functional theory (DFT) level using PBE96 functionals with the code WIEN2k. [32] GGA+U calculations were also performed to examine the effect of interactions in Fe 3d electrons, with effective Coulomb parameters ($U$–$J$) varied in the range of previously reported values from 0 to 6 eV. [33–36]

### III. RESULTS

**A. Structural characterization**

Figure 2 shows the powder XRD patterns of the representative samples of undoped, Ca-doped, and F-doped LaOFeP. Almost all the diffraction peaks are assigned to those of the LaOFeP phase, indicating that the LaOFeP phase is dominant in all the samples, although there are several weak peaks attributable to FeP, LaP, and LaOF (indicated by arrows in the figure). Peak intensities of the impurity phases relative to those of LaOFeP are less than 3% for FeP, 4% for LaP, and 3% for LaOF, suggesting that the impurity contents are at these levels in all the samples. It should be noted that the detection limit of our XRD measurements is ~0.1 mol%, [37] which suggests that magnetic measurements can detect more trace impurities compared with those detected by the XRD measurements. In fact, our magnetization measurements indicated that a ferromagnetic impurity $Fe_2P$ [38] was often involved in the F-doped LaOFeP; however this was not detected in the XRD pattern. However, these impurities do not provide significant influence on the following superconductivity characterizations, because it has been confirmed that they do not show superconductivity down to 2 K.

The XRD measurements have shown that lattice parameters exhibit non-negligible scattering (i.e., larger than the experimental errors, $\Delta a$ = 0.00004 nm and $\Delta c$ = 0.0002 nm) even for the same nominal composition samples (the data will be shown in Fig. 9 in comparison with $T_c$). The undoped LaOFeP shows the scattering with $a$-axis lengths of $a$ = 0.3962 ± 0.0001 nm (the latter figure shows the scattering range of the $a$-axis length) and $c$-axis lengths of $c$ = 0.8511 ± 0.0002 nm. The Ca-doped samples have similar $a$-axis lengths $a$ = 0.3961 ± 0.0001 nm, while c-axis lengths are $c$ = 0.8516 ± 0.0004 nm, which are larger than those of the undoped LaOFeP. The F-doped samples have much smaller $a$-axis lengths and the smallest $c$-axis lengths, with the scattering in the $c$-axis length ($a$ = 0.3957 ± 0.0001 nm, $c$ = 0.8503 ± 0.0004 nm). Such scatterings in the lattice parameters may be due to deviations in the chemical compositions, because it is difficult to eliminate a trace deviation from the stoichiometry for multinary compounds such as LaOFeP, even when synthesized in closed silica tubes. In spite of the scattering, there is a clear trend that the $a$-axis lengths of the Ca-doped and the F-doped LaOFeP, on average, are smaller compared with those of undoped samples by ~0.03 and ~0.13%, respectively. Since the ionic radius [39] of $Ca^{2+}$ (0.112 nm) is smaller than that of $La^{3+}$ (0.116 nm), and that of $F^-$ (0.131 nm) is smaller than that of $O^{2-}$ (0.138 nm) for the same coordination number, the shrinkage of the $a$-axis lengths in the Ca-doped and F-doped LaOFeP suggests that $Ca^{2+}$ and $F^-$ substitute $La^{3+}$ and $O^{2-}$, respectively.

**B. Electrical properties**

Figure 3 shows the temperature (T) dependences of the electrical resistivity ($\rho$) for the undoped (upper), Ca-doped (middle), and F-doped LaOFeP (lower) samples. Insets show expanded $\rho$–T curves in the temperature range below 15 K. The thick $\rho$–T curves correspond to those of the representative samples used for the magnetization measurements in the next section (note that these representative samples are employed in Figs. 2, 4, 5, 6, and 7). All the samples show metallic conduction in the normal conducting state. With decreasing temperature, sharp drops in $\rho$ to zero, which are more clearly visible in the insets, are observed at low temperatures. These drops are attributed to superconducting transitions, as confirmed by the magnetization measurements (Section III C). Superconducting transition temperatures ($T_c$) hereafter are defined as a temperature where $\rho$ takes a half value of the resistivity at 15 K ($\rho_{15K}$). It is observed that both the $T_c$ and $\rho_{15K}$ vary among the samples in a wide range. For instance, $T_c$ ranges from 2.4 to 5.5 K among the undoped samples, even though they were prepared in apparently the same conditions.

**C. Magnetic properties**

Figure 4 shows the temperature dependences of the magnetization ($M_{mol}$) of the representative samples of (a) undoped, (b) Ca-doped, and (c) F-doped LaOFeP measured at a magnetic field of 10,000 Oe. $M_{mol}$ gradually increases with decreasing temperature, and then sharply increases with a further decrease at temperatures below ~20 K. The sharp increase is attributable to the impure paramagnetic ions included in the La metal source. For instance, the sharp increase region follows the Curie paramagnetic equation $\chi_{mol} = C/T$ with $C = 1.1 \times 10^{-3}$ emu K/mol. The C value corresponds to ~0.14 mol% of $Ce^{3+}$, which can be understood as the involvement of 0.42 mol% Ce in the La metal source. Temperature dependences of the molar magnetic susceptibility (thicker lines ($\chi_{mol}$) in Figs. (a)-(c)) originating from the LaOFeP phase are extracted by subtracting the contributions of the Curie paramagnetic susceptibility. The $\chi_{mol}$ values of all the samples show a steep decrease at < 5 K.

In the normal conducting states (e.g., at >10 K), the $\chi_{mol}$ values are small and show weak temperature dependences, indicating that the samples contain no magnetic ions, or that the magnetic moments of the ions are very small. Therefore, paramagnetism of itinerant electrons is dominant in the normal conducting states, although Van Vleck paramagnetism due to orbital moments and the more negligible core electron diamagnetism may contribute to the observed susceptibility.

Figure 5 shows $M_{mol}$–H curves for the undoped, Ca-doped, and F-doped LaOFeP at 2 K. The inset shows an expanded view of the curves in the small H region. $M_{mol}$ decreases as H increases from 0 Oe and then increases when H exceeds threshold values (indicated by arrows in the insets of Fig. 5). The threshold values are regarded as lower superconducting critical fields ($H_{c1}$). The $H_{c1}$ values vary between 20 and 50 Oe and are comparable to those reported on other transition-metal-based superconductors having similar $T_c$ such as $Na_xCoO_2 \cdot H_2O$ and $Cu_{0.07}TiSe_4$ ($H_{c1}$ = 10–30 Oe). [40,41] The $M_{mol}$–H curves become almost linear at magnetic fields >10,000 Oe, indicating the gradual disappearance of diamagnetic shielding and paramagnetism becoming dominant at >$H_{c1}$. The slopes of the $M_{mol}$–H curves are −0.65, −1.48, and −3.08 for the undoped, Ca-doped, and F-doped samples, respectively, in a magnetic field region from zero to $H_{c1}$. These values are close to the perfect diamagnetic susceptibility (dashed line in the inset of Fig. 5), indicating that the negative magnetizations are due to the Meissner effect, and the volume fractions of the superconducting phases calculated from the slopes are 20%, 46%, and 97% for the undoped, Ca-doped, and F-doped samples, respectively. These volume fractions, which are safely larger than those of the above impurity phases, substantiate the superconductivity being due to the LaOFeP phase. Magnetizations of the F-doped LaOFeP samples often exceed the perfect diamagnetization below $H_{c1}$, because demagnetization effects overestimate superconducting volume fractions in diamagnetic-shielding measurements. [42] The steep decreases of $\chi_{mol}$ values at ~5 K observed in Fig. 4 indicate the growth of superconducting volumes with decreasing temperature.

**D. Photoemission spectra**

**D-1. Fe $2p_{3/2, 1/2}$ core spectra: HXPES**

The chemical bonding nature of LaOFeP was examined from the chemical shifts of Fe $2p$ peaks in HXPES. The undoped LaOFeP was selected for this study to avoid effects of the dopant.

Figure 6 shows a HXPES spectrum of the representative undoped LaOFeP sample with binding energies ranging from 700 to 730 eV, where two peaks due to Fe $2p_{3/2}$ and $2p_{1/2}$ are observed. The binding energy of Fe $2p_{3/2}$ (~706.8 eV) is similar to those of Fe metal [43] and Fe–P inter-metallic compounds (FeP [44] and $LaFe_4P_{12}$ [45]). On the other hand, it is distinctly different from those of FeO [46] and $FeS_2$ [35], suggesting that the Fe–P chemical bond in LaOFeP has less of an ionic nature and is more likely to be metallic or covalent.

**D-2. Valence band spectra: HXPES and SXPES**

Figure 7 shows HXPES (a) and SXPES (c) spectra of the undoped LaOFeP with binding energies from 0 to 20 eV. The energy is measured from the Fermi energy. The Shirley method [47] is employed to remove the backgrounds, and their intensities are normalized with the most intense band D around 17 eV, which is assigned to La $5p_{3/2}$ and $5p_{1/2}$, as discussed later. The Fermi edge is clearly observed both in the HXPES and SXPES spectra, which is consistent with the metallic conduction of LaOFeP. In addition to band D, the spectra show three distinct bands: A, B, and C. Band A, which is at the Fermi edge, is much stronger in the SXPES spectrum compared with that in the HXPES spectrum, and band B is also enhanced in the SXPES spectrum compared with that in the HXPES spectrum. On the other hand, the intensities of band C are similar to each other in HXPES and SXPES, although its peak position in the HXPES spectrum is at a larger binding energy compared with that in the SXPES spectrum. The assignment of these bands is discussed with the results of band calculations in Section IV C.

## IV. DISCUSSION

**A. Correlation of transition temperature with crystal structure**

As mentioned in Section III B, we observed that both the $T_c$ and the $\rho_{15K}$ of undoped LaOFeP vary widely, e.g., $T_c$ ranges from 2.4 to 5.5 K even though the preparation conditions were apparently the same (Fig. 3). Further, the doping of Ca and F ions enhances $T_c$. Here, we discuss the possible parameters, such as the carrier density at the Fermi level ($D_{Ef}$) and lattice parameters, which control $T_c$.

As suggested by the theory of metals, conductivity is expected to be proportional to $D_{Ef}$, provided that the carrier mobility remains unchanged. However, it is difficult to obtain $D_{Ef}$ values experimentally, and therefore, $\rho_{15K}$ values in Fig. 3 are employed as a measure of $D_{Ef}$ with a supposition that the carrier mobility does not change with the chemical composition in LaOFeP. The $T_c$ values of the undoped, Ca-doped, and F-doped LaOFeP samples are plotted against $\rho_{15K}$ in Fig. 8. There is a good correlation between $T_c$ and $\rho_{15K}$ in the each sample with the same nominal compositions. However, $T_c$ decreases with increasing $\rho_{15K}$ in the undoped LaOFeP, but it increases in the doped LaOFeP samples. This result implies that the carrier density (and possibly $D_{Ef}$) is not the dominant factor for controlling $T_c$ in the case of the undoped and the aliovalent ion-doped LaOFeP samples.

Figure 9 (a), (b), and (c) show the $T_c$ values of all the samples as functions of the lattice parameters $a$ and $c$, and the unit cell volume ($V$), respectively. $T_c$ shows a trend to increase with decreasing $c$ and $V$, while it is hard to find a systematic variation in $T_c$–$a$ curve. The changes in the lattice parameters in these LaOFeP samples would be due to deviations in the chemical compositions. Therefore, the $T_c$ variation is also related to the compositional deviation. On the other hand, in addition to the non-stoichiometry, the doping of the aliovalent ions shows a different trend in the $T_c$–lattice parameters compared to the undoped sample. This may reflect the fact that the variations in the carrier density (i.e., $D_{Ef}$) with the lattice parameters differ between the aliovalent ions-doped LaOFeP and the undoped LaOFeP. In other words, 'chemical pressure' would have different effects on $T_c$ in these LaOFeP. It should be noted that the $T_c$ is more widely varied with the lattice parameters in the undoped samples than in the doped ones.

**B. Valence band structure**

HXPES core spectrum (Fig. 6) suggests that the Fe–P chemical bond is more likely to be covalent (or metallic). Here, we discuss the electronic states at the Fermi surface in more detail from the valence band photoemission spectra (Fig. 7). It is known that HXPES has a high sensitivity to an *s*-electron, a moderate sensitivity to a *p*-electron, and a low sensitivity to a *d*-electron, [48] whereas SXPES exhibits an opposite tendency. Therefore, the stronger intensities of bands A and B in the SXPES spectrum compared with the same in the HXPES spectrum in Fig. 7 suggest that these bands have larger portions of *p*- or *d*-electrons. In addition, the intensity of band A relative to that of band B is larger for SXPES, suggesting that the *d*-electron states are incorporated in band A. This suggestion, together with the expectation that the FeP layer acts as a conduction path, tentatively allows us to conclude that the Fermi surface is composed of the hybridized orbitals of Fe 3*d* and P 3*p*.

**C. Calculated band structure**

To further explore the Fermi surface, we performed spin-polarized first-principles energy band calculations to obtain the projected density of states (PDOS). It should be noted that the self-consistent field (SCF) cycles performed in these calculations converged to two similar but different states for LaOFeP. Therefore, strict convergence criteria (0.00001 Ry for energy convergence and 0.0001 electrons for charge convergence) were adopted, and the ground state was judged from the SCF total energy. [49] Figure 10 shows the PDOS calculated for ideal LaOFeP crystal with $U–J = 0$. The calculated total DOS in Fig. 10 reproduces well the experimental PES spectra in Fig. 7, although the calculated DOS peaks are inevitably shallower than the PES peaks. [50] It shows that LaOFeP has no energy gap showing metallic nature. It also shows that the $U–J = 0$ eV band structure of LaOFeP is slightly spin-polarized with a spin moment of 0.11 $\mu_B$ per Fe ion. It should be noted that there is a striking difference in calculated spin moment per Fe ion between LaOFeP (0.11$\mu_B$) and LaOFeAs (2$\mu_B$). [51] To confirm that this result was not affected by incorporation of electron correlation, GGA+U calculations were performed with the $U–J$ values varied up to 6 eV for Fe 3*d* electrons. These calculations provided similar spin-polarized metallic bands, even though the details of the PDOSs were different and the spin moments increased from 0.11 to 0.63 $\mu_B$ per Fe ion with increasing *U*–*J* value. This result confirms the present conclusion (i.e., the slightly spin-polarized metallic state) does not depend on electron–electron correlation in Fe 3*d* electrons. As noted in Fig. 4, the magnetic measurements indicate that the magnetic moments are almost quenched in LaOFeP. However, we observed paramagnetic behavior, and the paramagnetic behavior were weakly temperature dependent. It is difficult to discriminate the observed behavior from the small magnetic moments suggested in the calculations, and therefore the calculation results do not contradict the observations. Hereafter, we will employ the up-spin PDOS with *U*–*J* = 0 for comparison with the experimental PES spectra, because the selection of the *U*–*J* parameter does not change the PDOS structure to a great extent and the polarization is small, as explained above. The up-spin PDOSs (Fig. 11) show that Fe 3*d* orbitals widely spread in the valence band region (note that the Fermi energy is at 0 eV), and they overlap with P 3*p* orbitals. In addition, Fe 3*d* orbitals have non-negligible hybridization with P 3*s* [44] around 11 eV, whose energy corresponds to band C in Figs. 7 (a) and (c). In fact, the intensities of band C are mainly due to P 3*s* in HXPES and Fe 3*d* in SXPES, which leads to the shift in the peak position discussed in Section III D-2. The calculated PDOSs also show that band A is assigned to the states composed of Fe 3*d* and P 3*p*, band B to those of O 2*p* and P 3*p*, and band D to those of La 5*p* and O 2*s*. Figures 7 (b) and (d) illustrate the PES spectra schematically by considering that the photoelectron yields are larger for SXPES compared with that for HXPES by the ratio of 1:10:60 for *s*:*p*:*d* electrons, respectively.

Figure 12 shows the up-spin PDOSs decomposed to irreducible expressions for the atomic orbitals in the Muffin-Tin spheres defined for the L/APW+lo calculation. Because the Fe site in LaOFeP has point group symmetry of $D_{2d}$, the Fe 3*d* levels are split into three non-degenerated, $A_1$ ($d_{z^2}$), $B_1$ ($d_{x^2-y^2}$), and $B_2$ ($d_{xy}$), and a doubly degenerated E ($d_{xz}$, $d_{yz}$) representations. It is clearly observed in Fig. 12 (a) that the distribution of P 3*p* is similar to that of Fe 4*s*, although Fe 4*s* shows significantly smaller PDOS compared with that of P 3$p$. Similarly, Fe 4$p$ shows significantly smaller PDOS compared with that of P 3$p$. Regardless of the small contributions, these facts suggest that P 3$p$ and Fe 4$s$ form bonding orbitals. The P 3$p$ levels are split into two irreducible representations, i.e., a non-degenerated $p_z$ and a doubly degenerated ($p_x$, $p_y$), in a tetragonal lattice. As shown in Fig. 12 (a), P 3$p_z$ has non-negligible contributions around the Fermi energy, whereas the contribution of 3($p_x$, $p_y$) is negligible, indicating that P 3$p_z$ contributes mainly to the Fermi surface. It is also shown that the Fe 3$d$ levels further split into bonding and antibonding levels by forming hybridized orbitals with ligand P 3$p$ orbitals (representative peak positions are indicated by vertical bars in Fig. 12 (b)). This result shows that the Fe 3$d$ levels are roughly classified into two groups: the higher-energy (3$d_{x2-y2}$ and 3$d_{z2}$) and the lower-energy (3$d_{xy}$ and 3[$d_{xz}$, $d_{yz}$]) group. The inset of Fig. 12 (b) shows that the Fermi surface is composed of P 3$p_z$, Fe 3$d_{z2}$, and Fe 3($d_{xz}$, $d_{yz}$), indicating that these orbitals form the carrier transport paths and are responsible for the superconductivity. Figure 13 illustrates a simplified energy diagram built from these results.

We now comment on the quantitative value of the density of state at $E_f$ ($D_{Ef}$) in LaOFeP. The L/APW+lo calculation gave us the $D_{Ef}$ value of 3.4 states/eV per unit cell. On the other hand, when the Van Vleck paramagnetism and core state diamagnetism can be neglected, assuming that paramagnetic susceptibility can be explained by $\chi_{Pauli} = \chi_0(1+cT^2)$, [52] where $c$ is a constant and $\chi_0$ is the susceptibility at zero temperature, an experimental $D_{Ef}$ estimated from the $\chi_{mol}$ value of undoped LaOFeP at 20 K was 18 states/eV per unit cell. This value is six times larger compared with that of the L/APW+lo calculation. This enhancement is consistent with the result obtained by the heat capacity measurements. [53] Such an enhancement suggests that spin fluctuation dominates the magnetic properties of the present materials. [54]

## V. CONCLUSIONS

1. LaOFeP is a superconductor with a transition temperature ($T_c$) of 2.4–5.5 K. The observed $T_c$ variation is presumably due to changes in the lattice parameters and /or carrier density caused by the compositional deviation from stoichiometry. Aliovalent ion doping of Ca or F increases the value of $T_c$ up to 7 K. A linear correlation was found between the $T_c$-increase and unit cell volume; however, the increase in the undoped samples is larger compared with that in the doped samples.

2. The temperature-dependent magnetization indicates that the magnetic moment in LaOFeP is almost quenched, while spin-polarized first-principles energy band calculations suggest the presence of small (~0.11 $\mu_B$ /Fe) but finite spin moments at the ground state. This is compatible with the experimental results suggesting that LaOFeP exhibits a paramagnetic behavior with the weakly temperature-dependent $\chi_{mol}$ in the normal conducting state.

3. The photoemission spectra, together with the calculated PDOSs, clarify that the Fermi surface in LaOFeP is primarily constituted from Fe 3d orbital. The participation of P 3p orbital hybridizing with Fe 3d orbitals remains ~ 10%. These results verify that the present material is a Fe-based superconductor. Fe 3$d$ electrons, which are itinerant in the overlapped orbitals of Fe 3$d_{z2}$, Fe 3($d_{xz}$, $d_{yz}$) and P 3$p_z$, are responsible for the superconductivity and paramagnetism.

**Acknowledgements**

We are indebted to Drs. Yoshio Kobayashi, Sigenori Ueda, Jung-Jin Kim, Masaaki Kobata, Shin-ichi Fujimori, Yoshinori Nishino, Kenji Tamasaku, Makina Yabashi, Tetsuya Ishikawa, Kentaro Kayanuma, and Hidenori Hiramatsu for their help in the measurements and their fruitful discussions. The synchrotron radiation experiment (SXPES) was performed under the Common-Use Facility Program of JAEA (Proposal No. 2006B-E22).

**Figure captions**

FIG. 1. (Color online) Crystal structure of LaOFeP. The light-blue box represents the unit cell. $FeP_4$ tetrahedron units are shown as gray tetrahedra.

FIG. 2. (Color online) XRD patterns for the representative samples of undoped, Ca-doped, and F-doped LaOFeP. The vertical bars at the bottom represent the calculated positions of Bragg diffractions of LaOFeP. The arrows represent the diffraction peaks due to impurity phases.

FIG. 3. (Color online) Temperature (T) dependence of electrical resistivity ($\rho$) for undoped (a), Ca-doped (b), and F-doped LaOFeP (c) samples. Insets show expanded views below 15 K. Thick $\rho$ –T curves correspond to the representative samples.

FIG. 4. (Color online) Molar magnetization ($M_{mol}$) and extracted molar susceptibility ($\chi_{mol}$) of the representative samples of undoped (a), Ca-doped (b), and F-doped LaOFeP (c) as a function of temperature (T) measured in a magnetic field (H) of 10,000 Oe. Steep increases at T<20 K are due to paramagnetic impurities (e.g., $Ce^{3+}$). Curie paramagnetic susceptibilities are calculated as C/T, (C = $1.1 \times 10^{-3}$ emu K/mol, ~0.14 mol% of $Ce^{3+}$). The $\chi_{mol}$ curves are obtained by subtracting the Curie paramagnetic susceptibilities from the measured $M_{mol}$/H curves.

FIG. 5. (Color online) Molar magnetization ($M_{mol}$) versus magnetic field (H) for the representative samples of undoped, Ca-doped, and F-doped LaOFeP measured at 2 K. The inset shows expanded $M_{mol}$–H curves near $H_{c1}$ (indicated by arrows). The dashed line indicates the magnetization of the perfect diamagnetism ($M_{mol}$ = –3.20 H for Ca-doped and F-doped LaOFeP).

FIG. 6. (Color online) Fe $2p_{3/2}$ and $2p_{1/2}$ HXPES spectrum of undoped LaOFeP, together with reported X-ray photoemission spectroscopy (XPS) spectra of reference compounds. They are normalized with the Fe $2p_{3/2}$ intensity.

FIG. 7. (Color online) Valence band HXPES (a) and SXPES (c) spectra of the representative sample of undoped LaOFeP. They are normalized with the intensities of band D. (b) and (d) illustrate the contributions of atomic orbitals deduced from the calculated density of states.

FIG. 8. (Color online) Superconducting transition temperature ($T_c$) versus $\rho_{15\,K}$ for undoped (black circles), Ca-doped (red triangles), and F-doped (blue squares) LaOFeP. Solid lines are guides for the eyes.

FIG. 9. (Color online) Superconducting transition temperature ($T_c$) versus lattice parameters $a$ (a) and $c$ (b) and unit cell volume ($V$) (c) for undoped (black circles), Ca-doped (red triangles), and F-doped (blue squares) LaOFeP. Solid lines are guides for the eyes.

FIG. 10. (Color online) Densities of states in LaOFeP calculated with $U-J = 0$. The upper section is up-spin DOS and the bottom is down-spin DOS. The shadowed areas show the PDOS of Fe.

FIG. 11. (Color online) Up-spin PDOSs for each atom (La, O, P, and Fe) and total DOS in LaOFeP.

FIG. 12. (Color online) Up-spin PDOSs for P $3p$ and Fe $4s$, $4p$, $3d$ states in LaOFeP. The dashed line shows the Fermi energy. (a) PDOSs for P $3p$ and Fe $4s$, $4p$, $3d$. The shadowed areas indicate the PDOSs of P $3(p_x, p_y)$ and $p_z$ orbitals. (b) PDOSs for P $3p_z$ and Fe $3d_{x2-y2}$, $3d_{z2}$, $3d_{xy}$, $3(d_{xz}, d_{yz})$. Red bars indicate bonding (lower energy) and antibonding (higher energy) levels for Fe $3d$ – P $3p$

hybridized orbitals. Insets show expanded views, together with the PDOS for Fe 3*d* (black solid line) near the Fermi energy.

FIG. 13. (Color online) Simplified energy diagram of FeP layer in LaOFeP deduced from Fig. 12. (a) An extracted diagram for Fe 4*s*, Fe 4*p*, and P 3*p*. (b) An extracted diagram for Fe 3*d* and P 3*p*.

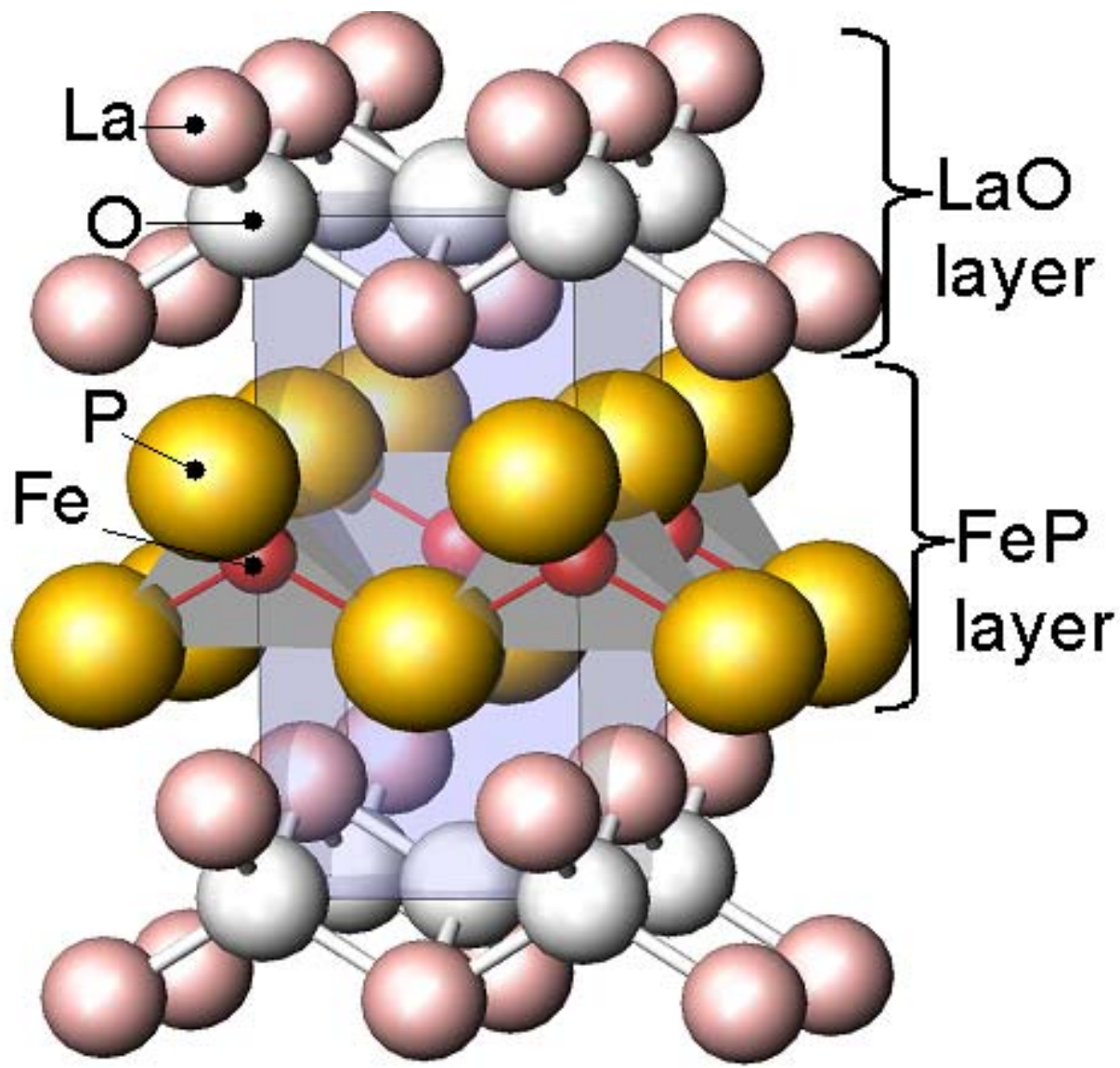

FIG. 1. Y. Kamihara, *et al*.

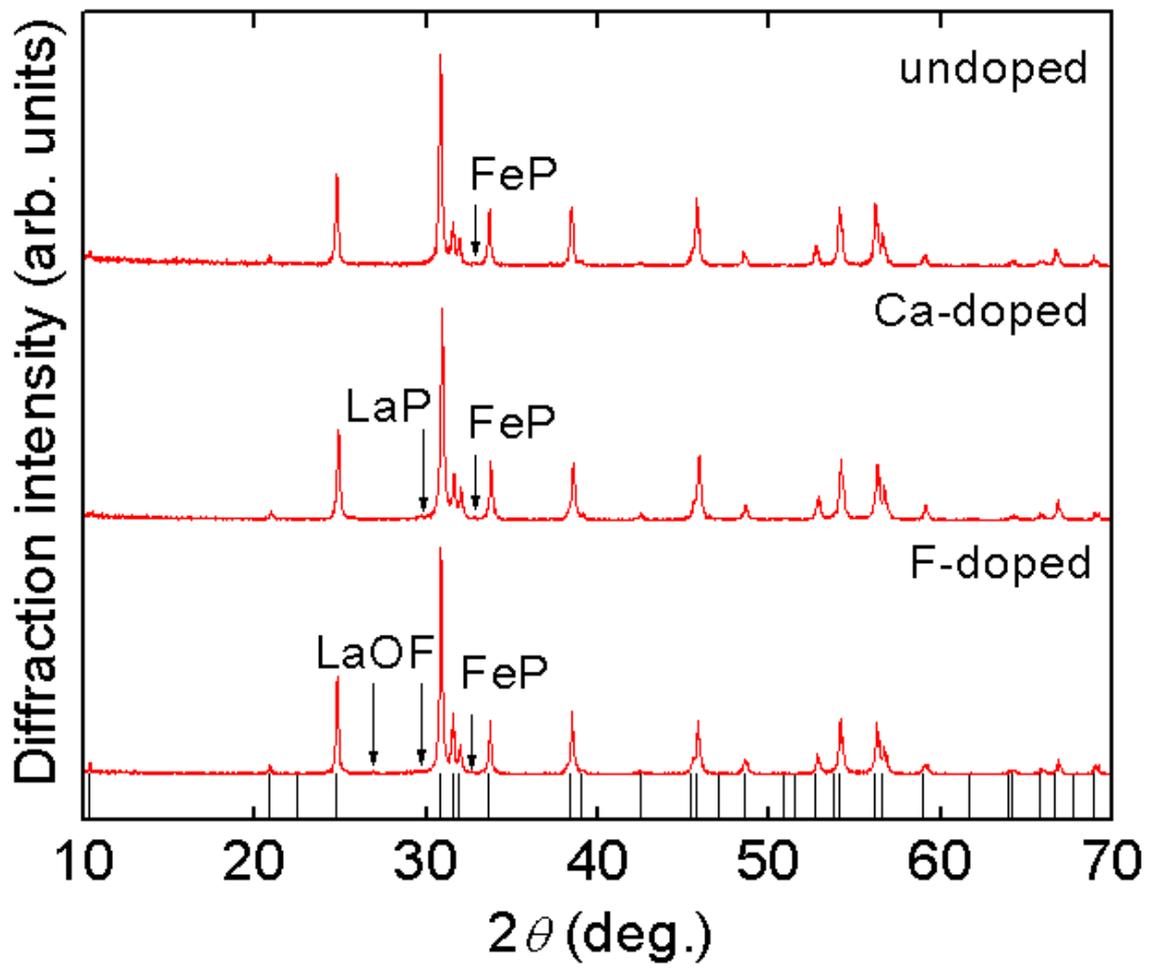

FIG. 2. Y. Kamihara, *et al*.

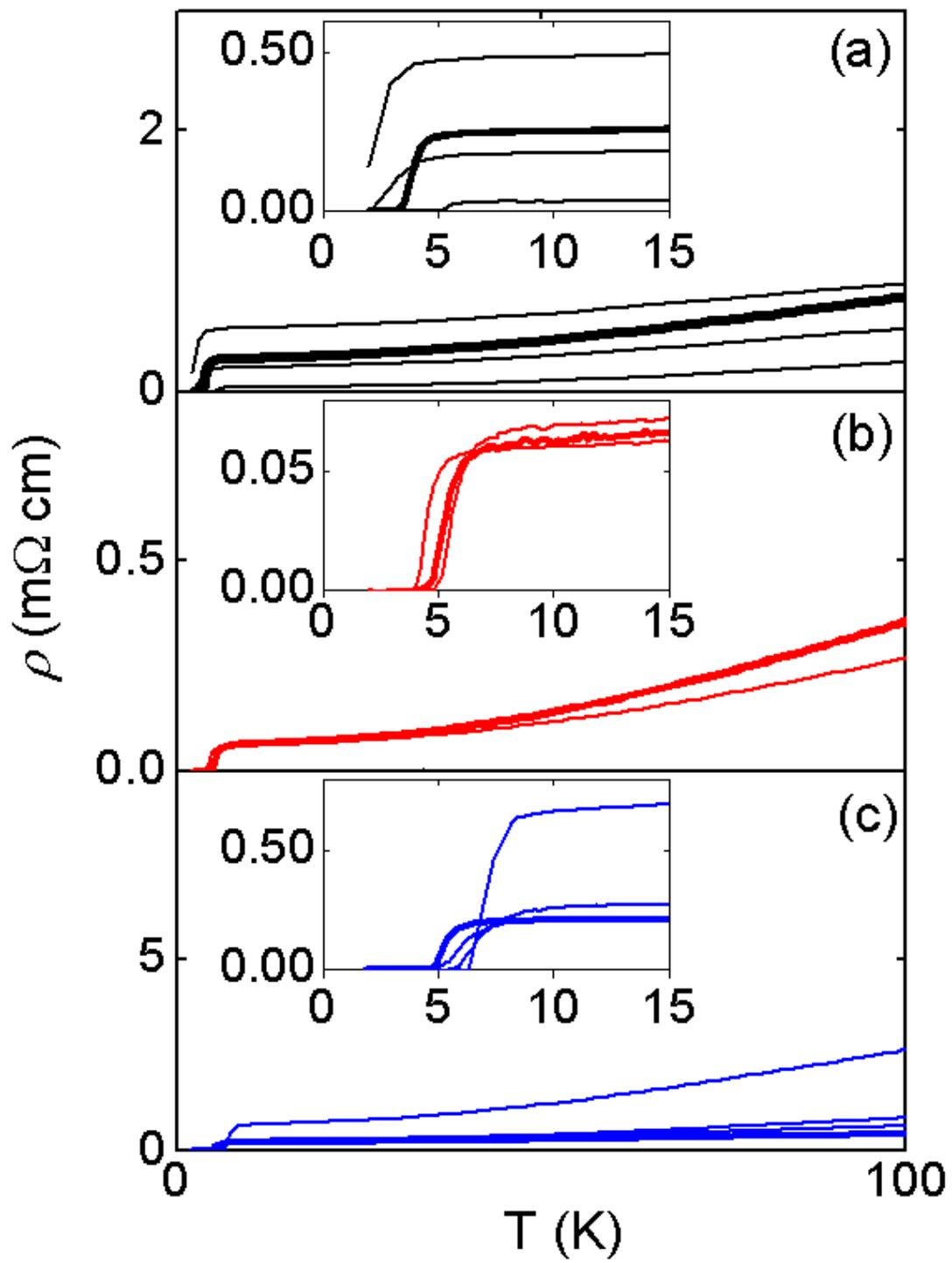

FIG. 3. Y. Kamihara, *et al*.

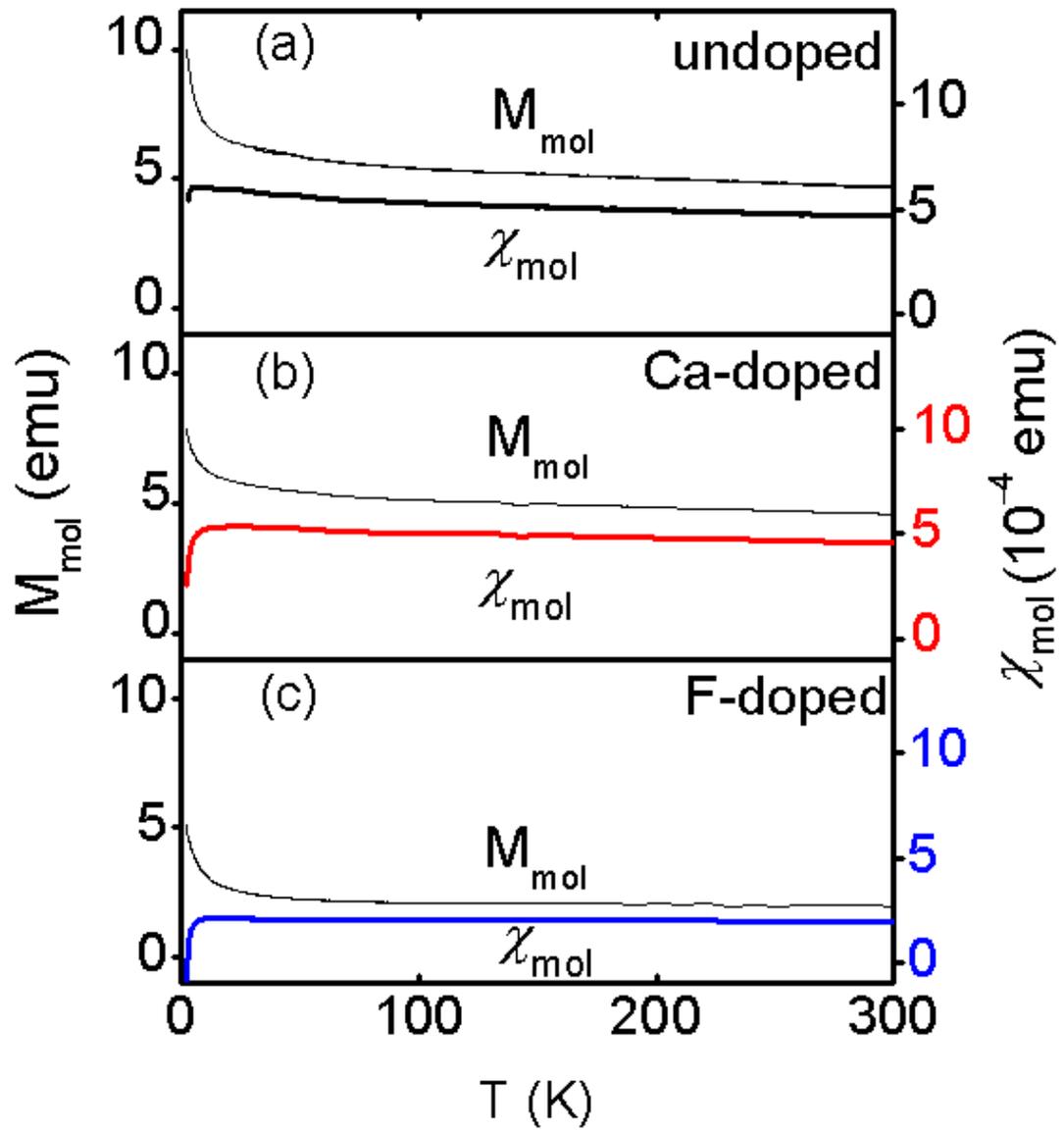

FIG. 4. Y. Kamihara, *et al*.

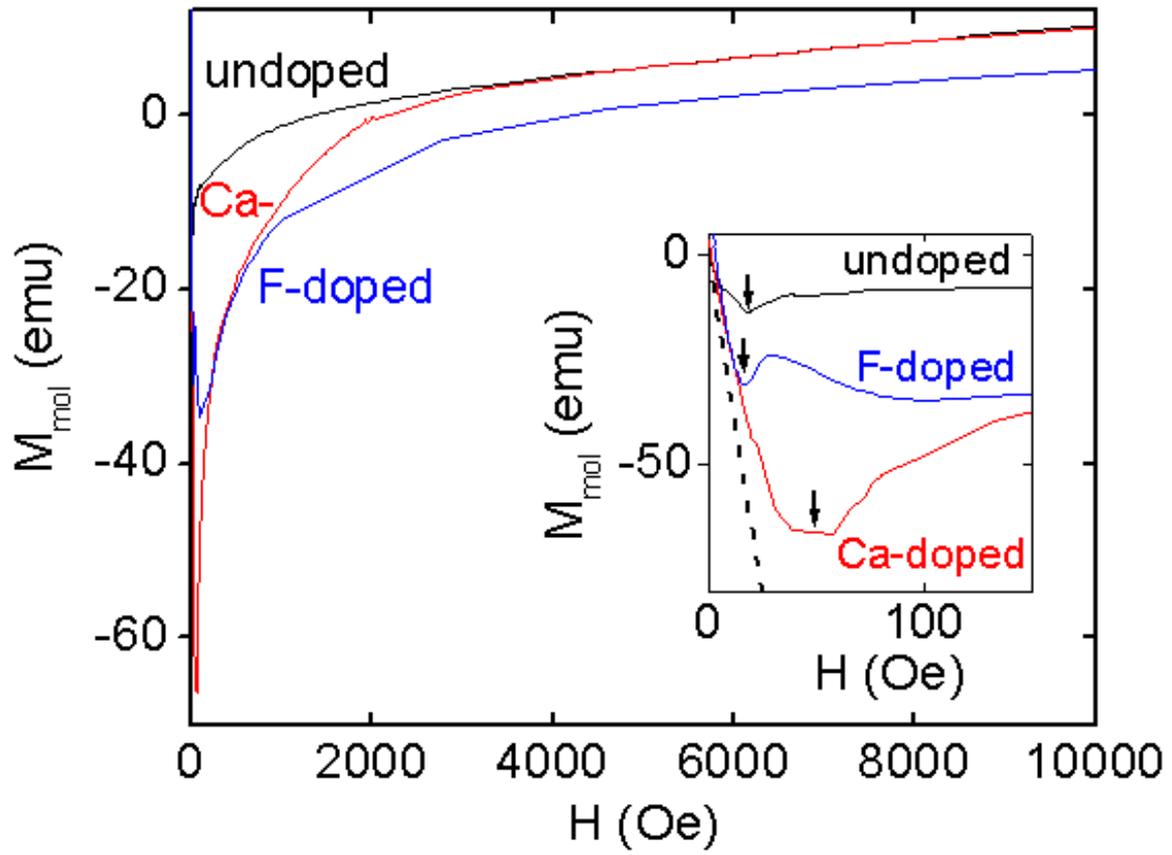

FIG. 5. Y. Kamihara, *et al*.

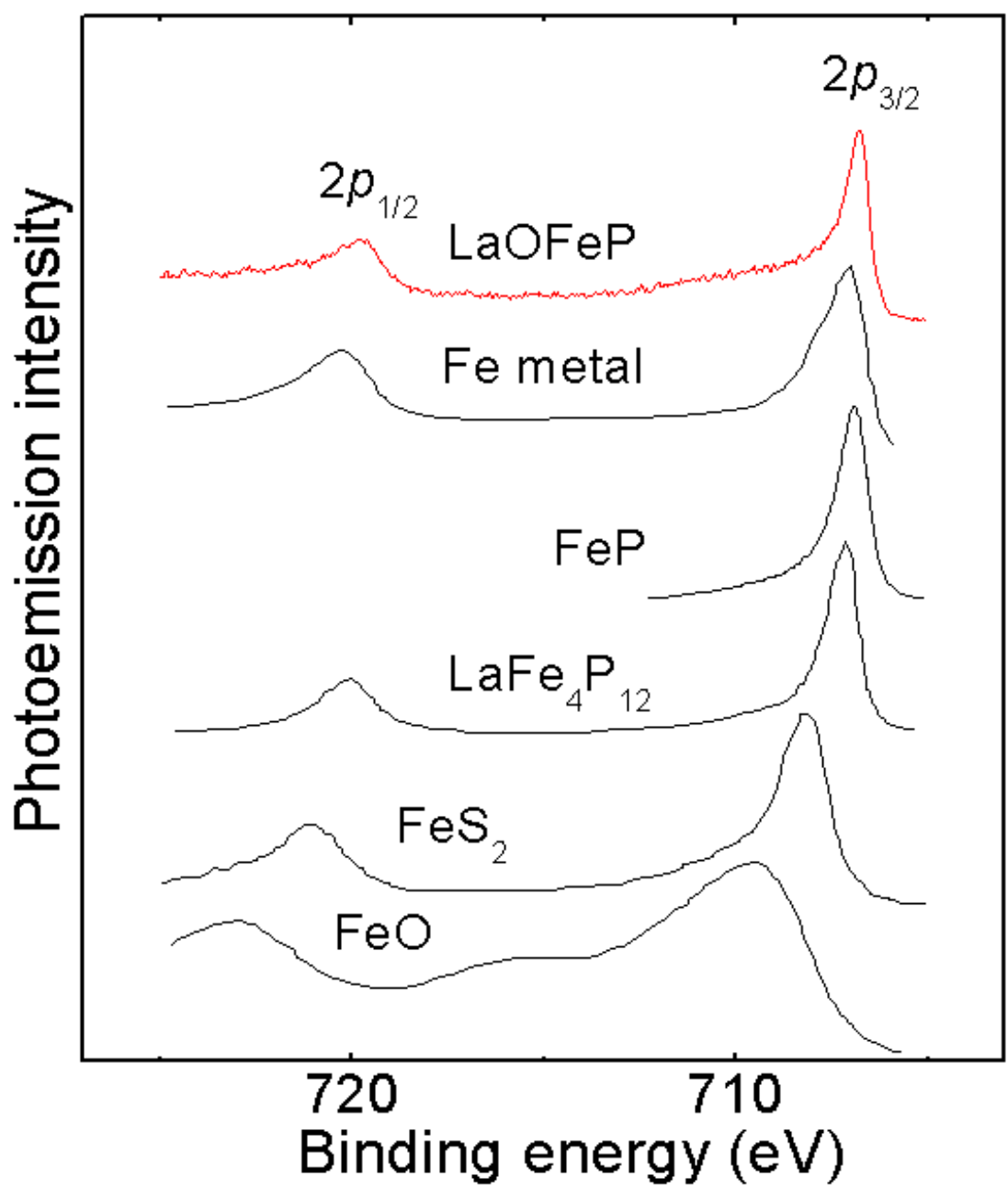

FIG. 6. Y. Kamihara, *et al*.

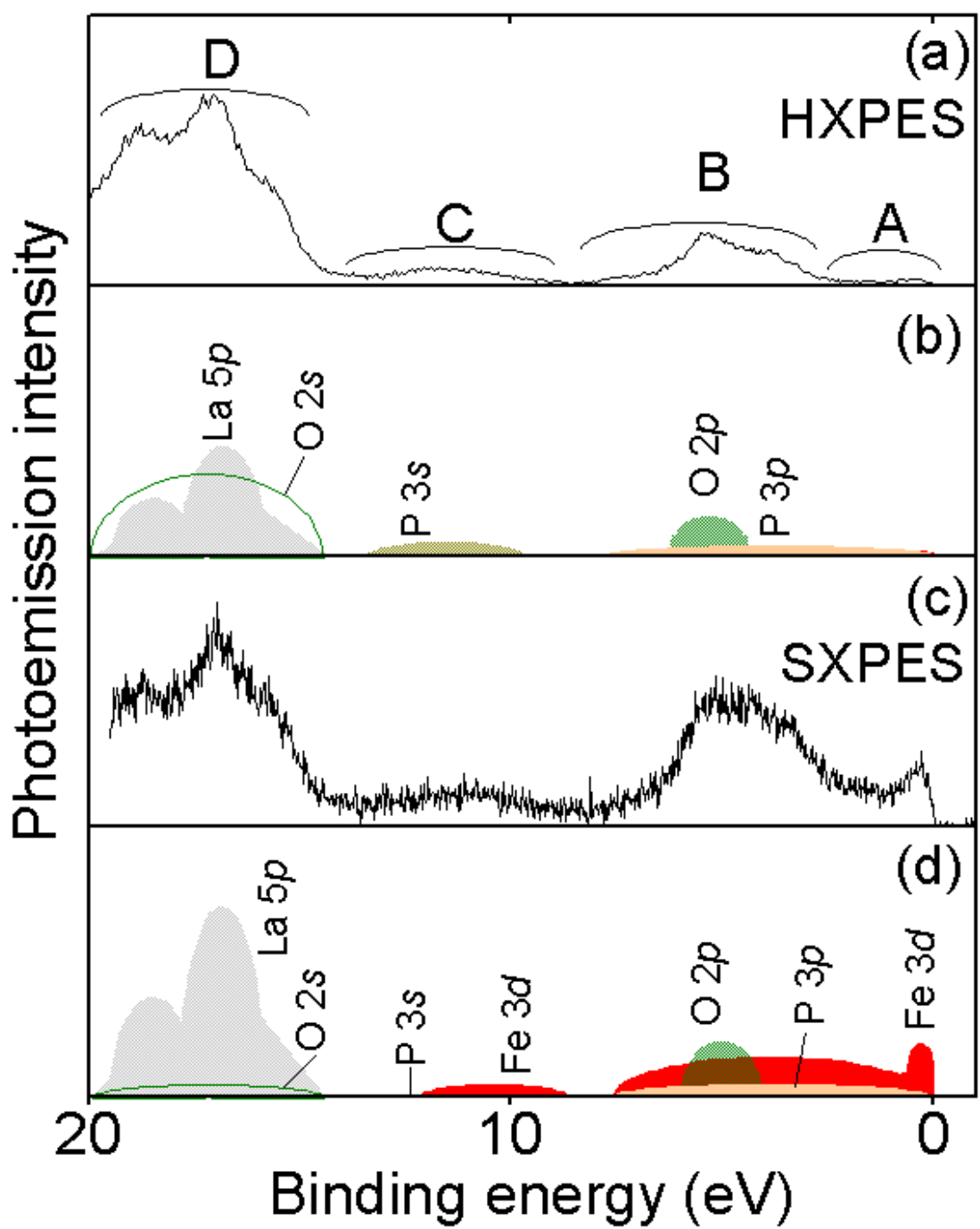

FIG. 7. Y. Kamihara, *et al*.

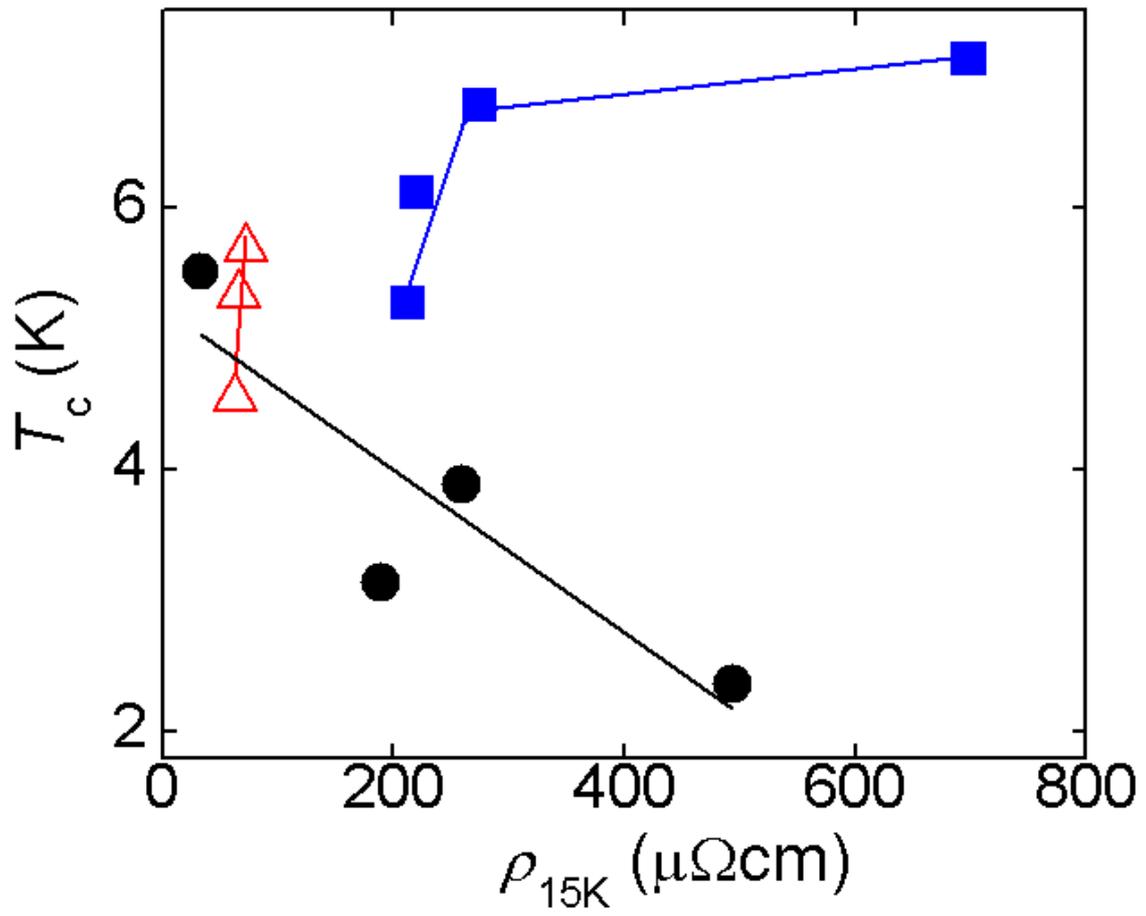

FIG. 8. Y. Kamihara, *et al*.

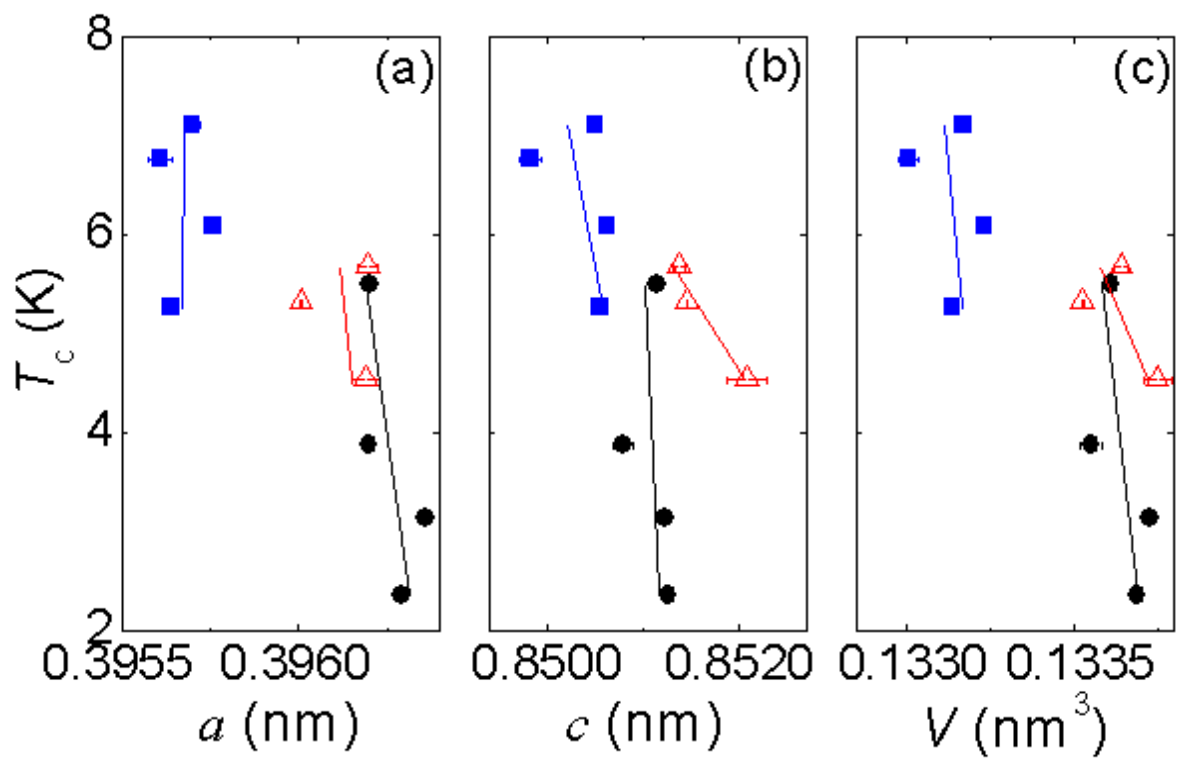

FIG. 9. Y. Kamihara, *et al*.

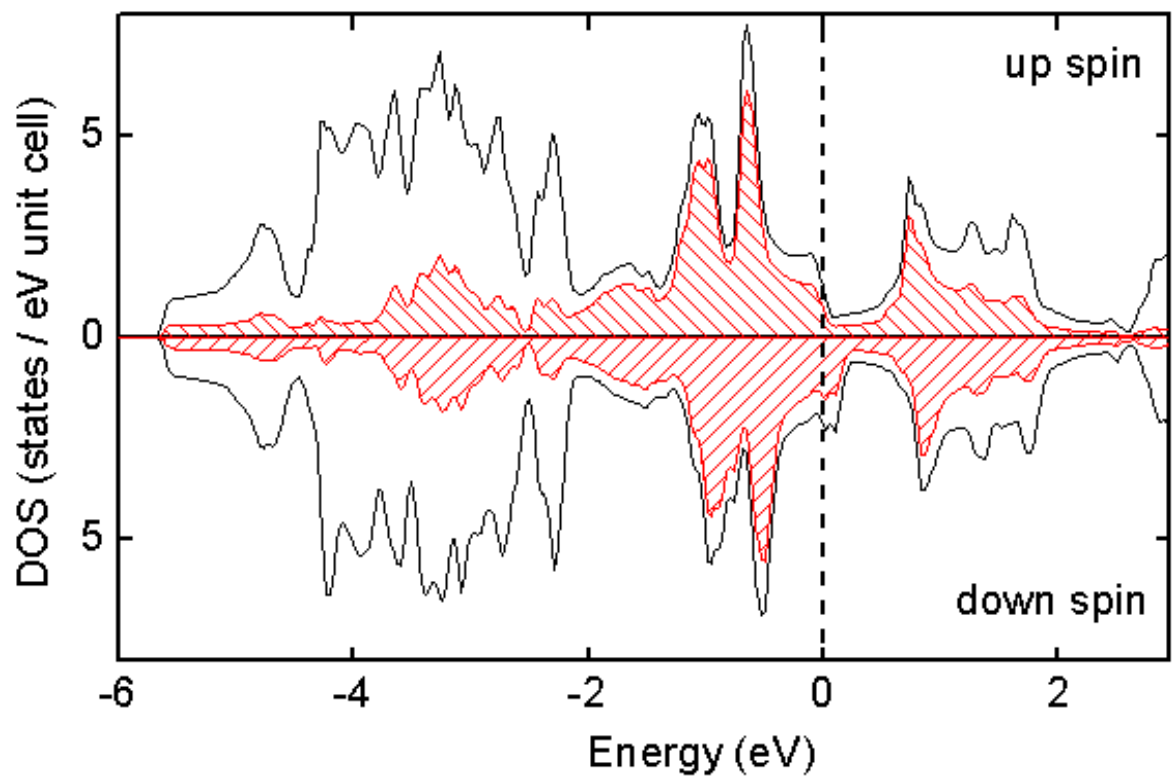

FIG. 10. Y. Kamihara, *et al*.

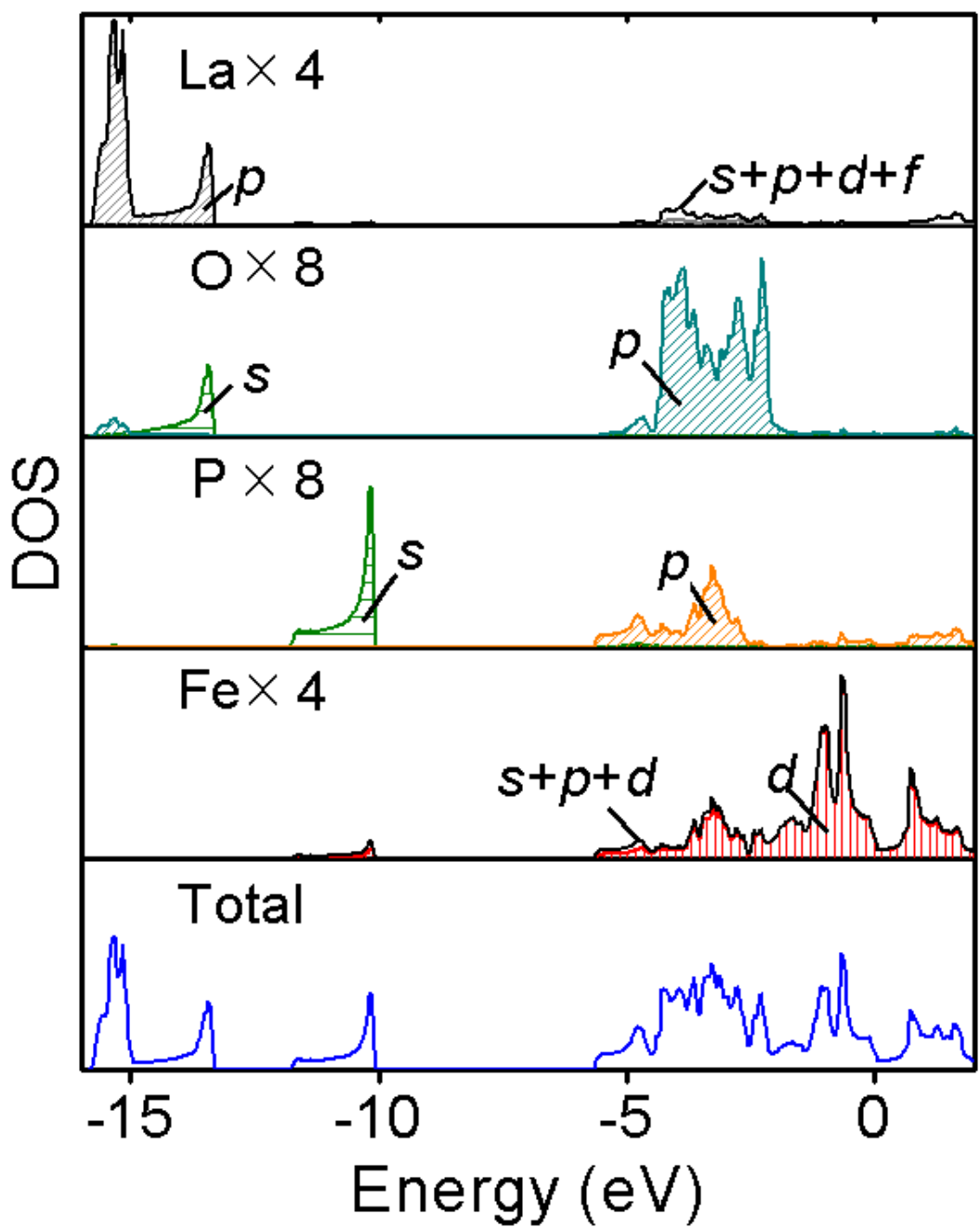

FIG. 11. Y. Kamihara, *et al.*

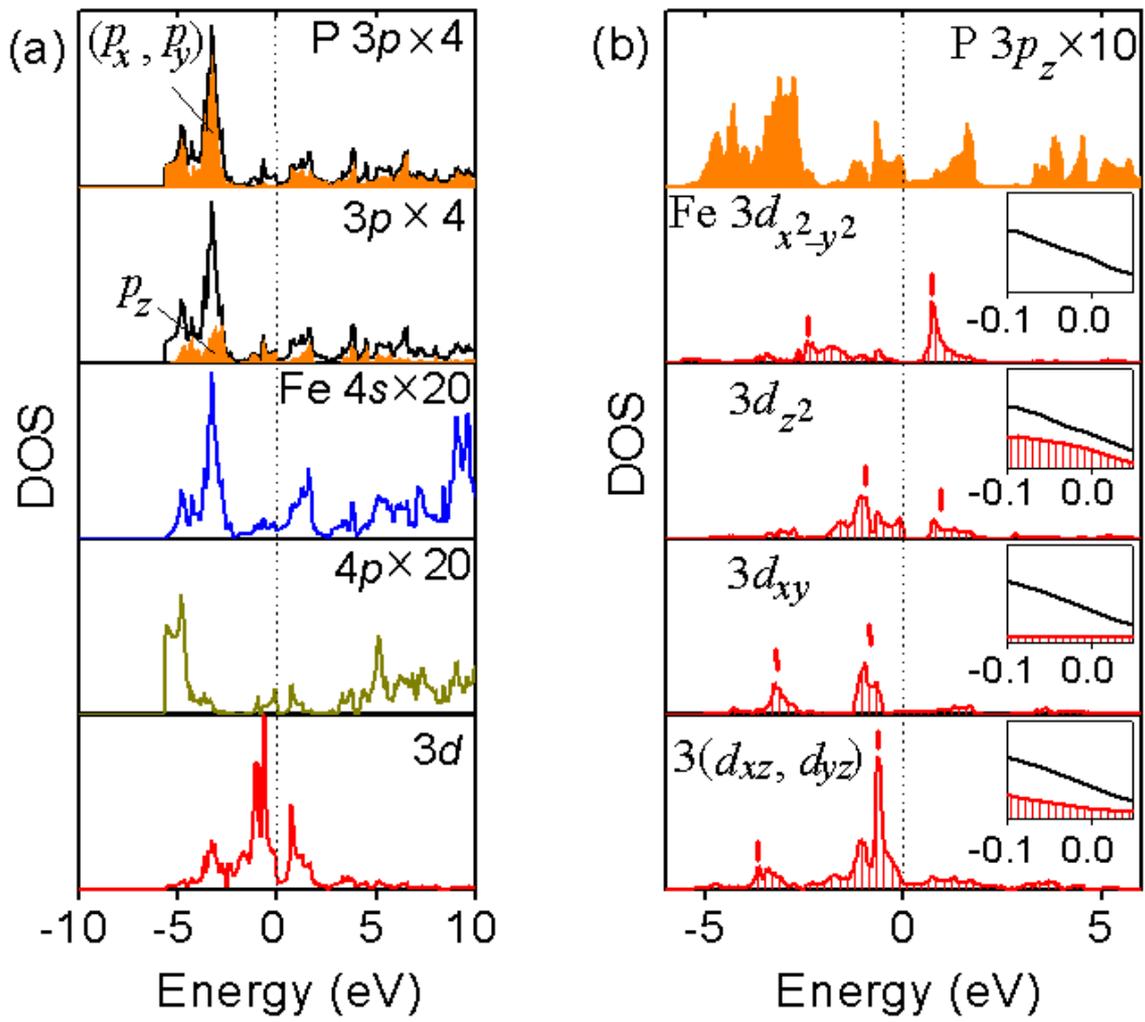

FIG. 12. Y. Kamihara, *et al*.

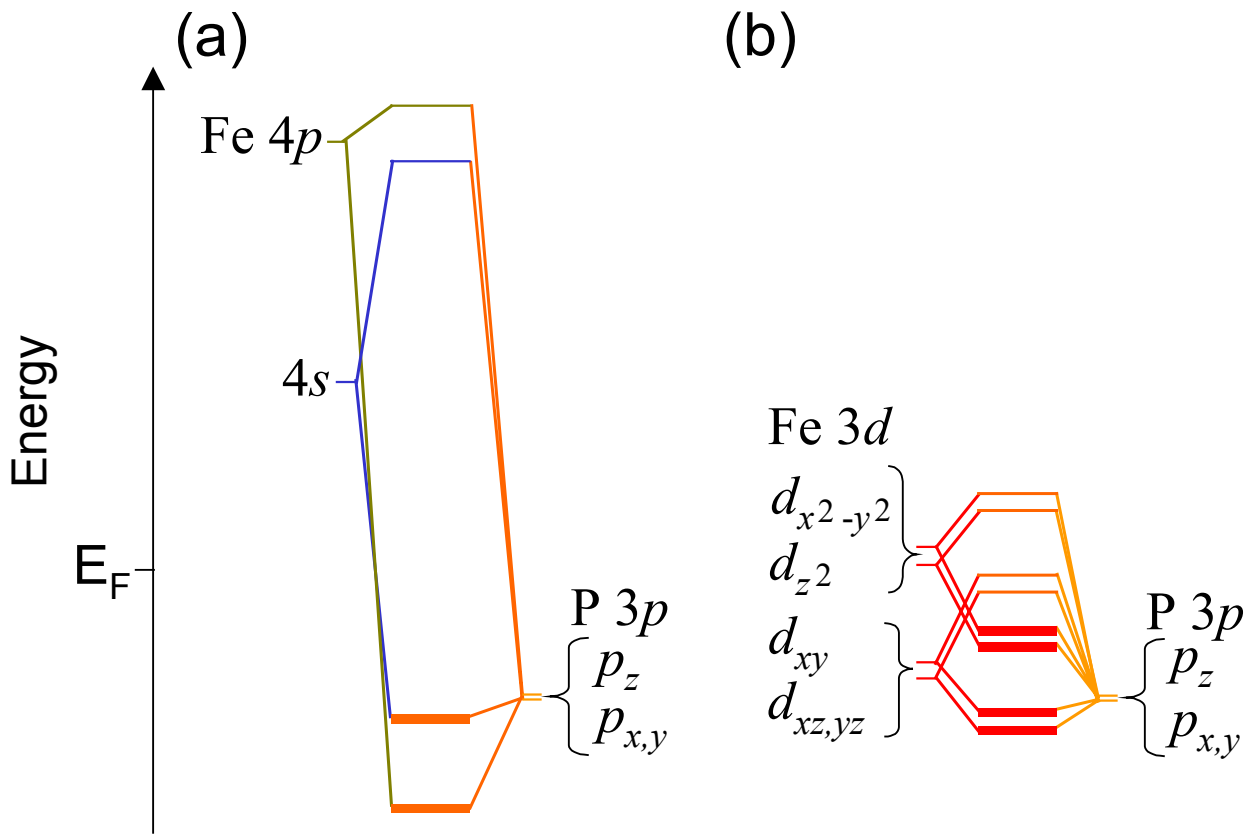

FIG. 13. Y. Kamihara, *et al*.